\documentclass[12pt]{article}
\usepackage{a4wide}
\usepackage{epsfig}

\def\half{{1\over 2}}

\begin{document}

\title{Quantum Dynamics of the Slow Rollover Transition
in the Linear Delta Expansion}

\author{H.~F.~Jones\thanks{e-mail h.f.jones@ic.ac.uk}\,,
P.~Parkin\thanks{e-mail p.parkin@ic.ac.uk}\, and D.~Winder\thanks
{e-mail d.winder@ic.ac.uk}\\ {\it Physics Department,
  Imperial College, London, SW7 2BZ, UK}}

\date{August, 2000}
\maketitle
\vspace{-3.5in}
\begin{flushright}
Imperial/TP/99-0/35
\end{flushright}
\vspace{2.9in}
\begin{abstract}
We apply the linear delta expansion to the quantum mechanical
version of the slow rollover transition which is an important
feature of inflationary models of the early universe. The method,
which goes beyond the Gaussian approximation, gives results which
stay close to the exact solution for longer than previous methods.
It provides a promising basis for extension to a full field
theoretic treatment.
\end{abstract}


\section{Introduction}

Inflationary models of the early Universe rely on the slow
evolution of an inflaton field $\varphi$ from the initial unstable
vacuum state in which $\langle\varphi\rangle=0$ to the final
stable vacuum in which $\langle\varphi\rangle=\pm a$, say.
The effective potential $V_{\rm eff}(\varphi_c)$ giving rise to this
transition has the generic form of a gentle hill centred at
$\varphi_c=0$ with minima at $\varphi_c=\pm a$.

The transition can be discussed at various levels of
sophistication.  At the most na{\"\i}ve level one can think
classically in terms of a ball rolling slowly down the slope of the potential.
The corresponding quantum-mechanical problem, which is the subject of
the present paper, is the time-development of a state whose wave function
is initially concentrated around the position of the maximum of the
potential. The full treatment of the problem must, of course, be
formulated within the framework of quantum field theory.

The first treatment of the quantum mechanical problem was given by
Guth and Pi \cite{gpi}, who solved exactly the equation of motion
for an initial Gaussian wave-function in an upside-down harmonic
oscillator potential $V=-\half k x^2$.
This was followed by a paper by Cooper et al.~\cite{cpi}, who used
a Gaussian ansatz in the Dirac time-dependent variational principle
for the standard symmetry-breaking potential $V=\lambda(x^2-a^2)^2/24$. The
resulting ``Hartree-Fock" solution tracks the exact solution for
a short time, but departs from it before the time at which $\langle
x^2 \rangle$ reaches its first maximum. Several years later
Cheetham and Copeland~\cite{cc} went beyond the Gaussian approximation
by using an ansatz which included a second-order Hermite polynomial.
This represented an improvement on
the Hartree-Fock approximation, but still did not reproduce the
first maximum in $\langle x^2 \rangle$ of the exact wave-function.

In the present paper we tackle this problem afresh using the linear
delta expansion.  This is a method akin to perturbation theory, but
with the crucial difference that the form of the unperturbed
Hamiltonian $H_0$ is not fixed once and for all, but varied at each
order in the expansion by some well-defined criterion. The role of
the formal parameter $\delta$ is simply to keep track of the order of
the expansion. The method has the great advantage that its generalization
to field theory is straightforward. We are motivated to apply it
to this dynamical problem by its success with the static
properties of the anharmonic oscillator, where it can be proved
rigorously that it converges to the exact result when applied to the
finite-temperature partition function\cite{ad+hfj} and the energy
levels\cite{guida}.

The relevant Hamiltonian ($\hbar=1$) is
\begin{eqnarray}
H&=&-\half {\partial^2 \over \partial x^2} +
\lambda(x^2-a^2)^2/24 + const.\cr
&&\cr
&=& -\half {\partial^2 \over \partial x^2} -\half m^2 x^2 +
gx^4,
\end{eqnarray}
with $m^2=\lambda a^2/6$ and $g=\lambda/24$, which we split according to
\begin{equation}\label{split}
H= -\half {\partial^2 \over \partial x^2} \pm \half \mu^2 x^2 + \delta
g(x^4-\rho x^2),
\end{equation}
where $2g\rho=m^2 \mp \mu^2$. That is, we choose as bare mass term
$\pm \half\mu^2 x^2$. The sign of the term as well as the value of
$\mu$ will be determined as functions of $t$ after the perturbative
expansion has been carried out to a given order (at which stage
$\delta$ is set equal to 1) by the criterion of minimal sensitivity
(PMS) \cite{pms}, namely that
\begin{equation}\label{PMS}
{\partial \langle x^2 \rangle^\half \over \partial \mu} =0.
\end{equation}

Note that for either sign of the new mass term, $\mu$ has a
limited range. In case (i), when the mass term is $-\half\mu^2$,
we have $2g\rho=m^2-\mu^2$. The essence of the delta expansion is
that the extra term $-g \rho x^2$ in the interaction should
compensate as far as possible the original term $gx^4$, which
means that $\rho$ should be positive. Hence we require that $\mu^2
< m^2$. In case (ii), when the mass term is $+\half\mu^2$, the same
restriction will arise from the form of the zeroth-order solution
and the initial wave-function.

\section{Delta Expansion}

The two cases need to be treated separately. Which of them is
relevant at a given value of $t$ is determined by the PMS
criterion.

\subsection{Case (i)}
In this case the bare Hamiltonian is
\begin{equation}
H_0=-\half{\partial^2\over\partial x^2}-\half\mu^2 x^2
\end{equation}
It is useful to scale $x$ according to $x=y/{\surd\mu}$, so that
\begin{equation}
H_0=-{\mu\over 2}\left({\partial^2\over\partial y^2}+y^2\right).
\end{equation}
Given that the initial wave-function is a Gaussian of the form
$\psi(t=0)=A \exp(-B y^2)$, the zeroth-order equation of motion
$H_0\psi_0 = i\partial\psi_0/\partial t$ can be solved
exactly by a wave-function of the same form with $A$ and $B$
becoming functions of $t$. The equations they have to satisfy are
\begin{eqnarray}
i\dot B/\mu &=& 2B^2+\half,\cr
i\dot A/\mu &=& AB,
\end{eqnarray}
with solutions
\begin{eqnarray}
B&=&\half\tan(\eta_0-i\mu t)\cr
A&=& {\cal N}\left(\cos(\eta_0-i\mu t)\right)^{-\half}
\end{eqnarray}
where $\eta_0$ is determined by $B(t=0)=\half\tan\eta_0$ and the
normalization constant ${\cal N}$ by
$A(t=0)={\cal N}\left(\cos\eta_0\right)^{-\half}$.
This is precisely the solution of Guth and Pi for the upside-down
oscillator, with $m$ replaced by $\mu$.

To obtain a systematic perturbative expansion in powers of $\delta$ it
is useful to write $\psi=\varphi\exp(-B(t)y^2)$. The equation for $\varphi$
is then
\begin{equation}
i\dot\varphi=\mu\left[B\varphi+2By\varphi'-\half\varphi''+\delta\tilde{g}
(y^4-\tilde{\rho}y^2)\right],
\end{equation}
where we have scaled $g$ and $\rho$ according to $g=\mu^3\tilde{g}$
and $\rho=\tilde\rho/\mu$.

A general feature of perturbative expansions for a polynomial
potential of degree $p$ is that $\varphi$ is also a polynomial, of
degree $Np$ in $N$th order of the expansion. Thus in the present
case, expanding $\varphi$ as $\varphi=\sum \delta^n\varphi_n$, the first-order part
$\varphi_1$ is an (even) polynomial of degree 4, which we write as
$\varphi_1=a+by^2+cy^4$. The equations of motion for the coefficients
$a$, $b$ and $c$ are
\begin{eqnarray}\label{abc}
i\dot a/\mu &=&  Ba -b \cr
i\dot b/\mu &=& 5Bb-6c-\tilde{\rho}\tilde{g}A\cr
i\dot c/\mu &=& 9Bc+\tilde{g}A,
\end{eqnarray}
which can be solved successively in reverse order, using the
solutions for $A$ and $B$ previously determined. The initial
conditions at $t=0$ are $a=b=c=0$.
Thus $c$ is given by
\begin{equation}
c={-i\tilde{g}{\cal N}/8 \over (\cosh\tilde\theta)^{9/2}}\left\{
3\tilde\theta + 2\sinh 2\tilde\theta +{1\over 4}\sinh 4\tilde\theta
-c_0\right\},
\end{equation}
where $\tilde\theta=\mu t +i\eta_0$ and $c_0=3i\eta_0 +
2i\sin2\eta_0 +(i/4)\sin4\eta_0.$

Using this solution in the equation for $b$ we obtain
\begin{equation}
b={-i\tilde{g}{\cal N}\over (\cosh\tilde\theta)^{5/2}}\left\{
-\half\tilde\rho(\tilde\theta+\half\sinh{2\tilde\theta})
+b_0+{3i\over 4}\left[3\tilde\theta\tanh\tilde\theta
+\cosh^2\tilde\theta-c_0\tanh\tilde\theta\right]\right\},
\end{equation}
where
$$
b_0={i\over 2}\tilde\rho\eta_0+{i\over 4}\tilde\rho\sin{2\eta_0}
+{9i\over 4}\eta_0\tan\eta_0-{3\over 4}c_0\tan\eta_0-{3i\over
4}\cos^2\eta_0.
$$

Finally, using this solution in the equation for $a$ we obtain
\begin{eqnarray}
a={\tilde{g}{\cal N}\over (\cosh\tilde\theta)^{1/2}}
\,\Bigg\{
&-& \half\tilde\rho\tilde\theta\tanh\tilde\theta+b_0
\tanh\tilde\theta -a_0\cr &+& {3i\over
4}\left[3\left(-\half\tilde\theta{\rm sech}^2\tilde\theta
+\half\tanh\tilde\theta\right)+\tilde\theta +\half c_0{\rm
sech}^2\tilde\theta\right]\Bigg\},
\end{eqnarray}
where
$$
a_0=\half\tilde\rho\eta_0\tan\eta_0+ib_0\tan\eta_0 +{3\over
4}i\left[{3\over 2}\left(-i\eta_0{\rm sec}^2\eta_0+i\tan\eta_0
\right)+i\eta_0+\half c_0{\rm sec}^2\eta_0\right].
$$

\subsection{Case (ii)}

The zeroth-order equations in this case are
\begin{eqnarray}
i\dot B/\mu &=& 2B^2-\half,\cr
i\dot A/\mu &=& AB,
\end{eqnarray}
with solutions
\begin{eqnarray}
B&=& \half{\rm coth}(\eta_0+i\mu t) \cr
A&=&{\cal N}(\sinh(\eta_0+i\mu t))^{-\half}.
\end{eqnarray}
As mentioned in the Introduction, the restriction on $\mu$ in this
case comes from the form of $B(t=0)$ and the form of the initial
wave-function, which, in all the papers quoted, is taken as a
minimal wave-packet appropriate to a positive mass term $+\half m^2 x^2$.
In the present formulation this means that $B(t=0)=(1/2)(m/\mu)$.
But since $B(t=0)=(1/2){\rm coth}\eta_0<1/2$, we have the same
restriction on $\mu$, namely $\mu<m$, as in Case (i).

The first-order equations for $a$, $b$ and $c$ are identical in
form to Eq.~(\ref{abc}), but the driving terms $A$ and $B$ are now
different.

The solution for $c$ is now
\begin{equation}
c={-i\tilde{g}{\cal N}/8 \over (i\sin\tilde\theta)^{9/2}}\left\{
3\tilde\theta - 2\sin{2\tilde\theta}+{1\over 4}\sin{4\tilde\theta}
-c_0\right\}
\end{equation}
where $c_0=-3i\eta_0+2i\sinh{2\eta_0}-(i/4)\sinh{4\eta_0}$ and in
this case $\tilde\theta=\mu t -i\eta_0$.

Using this solution in the equation for $b$ we obtain
\begin{equation}
b={\tilde{g}{\cal N}\over (i\sin\tilde\theta)^{5/2}}\left\{
-{i\over 2}\tilde\rho(\tilde\theta-\half\sin{2\tilde\theta})
+b_0-{3\over 4}\left[-3\tilde\theta{\rm cot}\tilde\theta
+\cos^2\tilde\theta+c_0{\rm cot}\tilde\theta\right]\right\},
\end{equation}
where
$$
b_0=\half\tilde\rho\eta_0-{1\over 4}\tilde\rho\sinh{2\eta_0}
-{9\over 4}\eta_0\coth\eta_0+{3i\over 4}c_0{\rm coth}\eta_0
+{3\over 4}\cosh^2\eta_0.
$$
Finally, using this solution in the equation for $a$ we obtain
\begin{eqnarray}
a={\tilde{g}{\cal N}\over (i\sin\tilde\theta)^{1/2}}
\,\Bigg\{
& &\half\tilde\rho\tilde\theta{\rm cot}\tilde\theta+ib_0
{\rm cot}\tilde\theta +a_0\cr &-& {3i\over
4}\left[{3\over 2}\tilde\theta{\rm cosec}^2\tilde\theta
+\half{\rm cot}\tilde\theta-\tilde\theta -\half c_0{\rm
cosec}^2\tilde\theta\right]\Bigg\},
\end{eqnarray}
where
$$
a_0=\half\tilde\rho\eta_0{\rm coth}\eta_0-b_0{\rm coth}\eta_0 -{3\over
4}\left[{3\over 2}\eta_0{\rm cosech}^2\eta_0+\half{\rm coth}\eta_0
+\eta_0-{i\over 2} c_0{\rm cosech}^2\eta_0\right].
$$
We have checked these solutions by numerical integration using the
Runge-Kutta method. This reveals that in Case (ii) care needs to be
taken to ensure that we are on the appropriate branch of the square
roots. At values of $t$ where $\sin\tilde\theta=-1$, a na{\"\i}ve
numerical evaluation will stay on the first sheet, thus giving rise
to a discontinuity, whereas the true solution is, of course,
continuous.

In fact, as we shall see, the coefficient $a$ is not needed in the
calculation of $\langle x^2 \rangle^\half$ to first order in $\delta$,
though it would, of course, be needed in higher order.

\section{Variational Aspect}

The expressions we have obtained all depend on the parameter $\mu$
introduced in Eq.~(\ref{split}). The other essential aspect of the
delta expansion is that such a parameter is determined by some
non-perturbative criterion, most frequently the principle of
minimal sensitivity, Eq.~(\ref{PMS}).

To that end we need an expression for $\langle x^2 \rangle$,
which, given that the wave-function is a (complex) Gaussian with
polynomial corrections, can be written down in closed form in terms
of the coefficients $A, B, a, b, c$. Thus to order $\delta$,
\begin{equation}
|\psi|^2 = \left[|A|^2 + 2\delta{\rm
Re}\left\{A^*(a+by^2+cy^4)\right\}\right] {\rm e}^{-\alpha y^2},
\end{equation}
where $\alpha=2{\rm Re} B$, so that
\begin{eqnarray}\label{rms}
\langle y^2 \rangle &=& {1 \over 2\alpha}
{1+(2\delta/|A^2|){\rm Re}\left\{A^*(a+3b/(2\alpha)+15c/(4\alpha^2))\right\} \over
 1+(2\delta/|A^2|){\rm Re}\left\{A^*(a+b/(2\alpha)+3c/(4\alpha^2))\right\} }\cr
 & & \cr
&=&{1 \over 2\alpha}\left[1+{2\delta\over |A^2|}{\rm Re}\left\{A^*({b\over
\alpha}+{3c \over \alpha^2})\right\}\right].
\end{eqnarray}
It is an interesting feature of the structure of the perturbative
equations that the wave-function is automatically normalized to the
order we are working. That is, $\int \varphi_0^*\varphi_1 = 0$. Thus the
second equation is identical to the first, and not merely an
O($\delta$) approximation to it. The expectation value we seek is
obtained on scaling by $\mu$, i.e. $\langle x^2
\rangle = \langle y^2 \rangle/\mu$.

At this stage we set $\delta=1$ and apply Eq.~(\ref{PMS}). This has to
be done for each time $t$, and the result is that the chosen value
$\bar\mu$ of $\mu$ now becomes a function of $t$, even though $\mu$
was treated as a constant in the equations of motion. In the
present case, since we are unable to go to very high orders in the
expansion this is a more important property than the $N$-dependence
of $\bar\mu$. The O($\delta^0$) calculation does not have such a
stationary point.

In Fig.~1 we show graphs of $\langle x^2 \rangle ^\half$ for
various values of $t$. The parameters chosen are those used in
Refs.~\cite{cpi} and \cite{cc}, namely $a=5$ and $\lambda=0.01$
(which corresponds to a ``large" dimensionless coupling
contant\cite{cpi}). We
include both cases by plotting $\langle x^2
\rangle^\half$ as a function of $\sigma\mu$, where $\sigma=-1$
for case (i) and +1 for case (ii). There is a well-defined maximum
which moves steadily to the right as $t$ increases, crossing over
from case (i) to case (ii) at about $t=11$. From these and similar
graphs we extract the value of $\bar\mu(t)$, which is plotted in Fig.~2.

Using these values of $\bar\mu(t)$ we can then calculate $\langle
x^2 \rangle^\half(\bar\mu(t),t)$ from Eq.~(\ref{rms}) as a function
of $t$. This is plotted in Fig.~3 along with the results obtained
using the ``Hartree-Fock" method of Ref.~\cite{cpi}, the improved
variational method of Ref.~\protect\cite{cc}, the exact value of $\langle
x^2 \rangle^\half(t)$, obtained by Fourier transform and numerical
integration\protect\cite{dm}, and finally the result of first-order
perturbation theory. The latter corresponds to O($\delta$) of the delta
expansion, but with $\mu$ fixed at $m$ in case (i), and exemplifies
the importance of the $t$-dependence of $\bar\mu$.

As can be seen, the delta-expansion result tracks the exact result
for longer than either of the other variational calculations,
essentially up to the point where $\langle x^2
\rangle^\half$ reaches its maximum, but then
overshoots. A similar degree of accuracy in quantum field
theory would mean that, to this order of the expansion,
the inflationary period would be
very well described, but the reheating process less
so. To extend the range of the approximation to longer times
a higher-order calculation would presumably be needed

\section{Discussion}

Figure 3 is our main result, but it is also of interest to enquire
how closely the calculated wave-function agrees with the exact
result, since a well-known feature of variational methods is that
quite reasonable values for expectation values such as $\langle x^2
\rangle$ can be obtained with rather inaccurate wave-functions.
In fact our wave-function agrees rather well with the true
wave-function up to $t\approx 6$, but begins to diverge from it thereafter,
even though still giving good values for $\langle x^2 \rangle$. In
Fig.~4 we plot the two values of $|\psi|^2$ versus $x$ for $t=6$ and 8.

Various extensions of the present treatment are possible. Given the
simplicity of the first-order equations resulting from the method
it seems that the extension to ${\rm O}(\delta^2)$ should be
relatively straightforward, certainly if the integrations are
performed numerically, and it would be interesting to see the
improvement thereby achieved. As mentioned earlier, the
wave-function would involve an even polynomial of order 8
multiplying the lowest-order Gaussian. Another possibility is the
use of the original delta expansion\cite{cmb}, whereby the $x^4$
term in the potential is written as $x^{2(1+\delta)}$ and expanded
as $x^2(1+\delta\ln x^2 + \dots)$. This expansion is known to
converge for the energy levels of the anharmonic oscillator, and
the ${\rm O}(\delta)$ calculation for the present problem should
be tractable. However, the disadvantage of this method is that its
extension to field theory beyond first order becomes extremely
difficult.

The most important extension is clearly to attempt to apply the
methodology of the linear delta expansion to the quantum field
theory problem. The importance of going beyond the Gaussian
approximation has been emphasized in refs.~\cite{cc} and
\cite{fernando}, and technically the linear delta expansion is
essentially a modified perturbation theory, modulo the
crucial variational aspect.

\section*{Acknowledgments}

We are grateful to F.~Lombardo for extremely useful discussions,
and D.~Monteoliva for making available the code for the exact
solution. P.~Parkin and D.~Winder gratefully acknowledge the financial
support of the Particle Physics and Astronomy Research Council.

\newpage

\begin{figure}[t]
\vspace{-1in}
\centerline{\epsfxsize=3in\epsffile{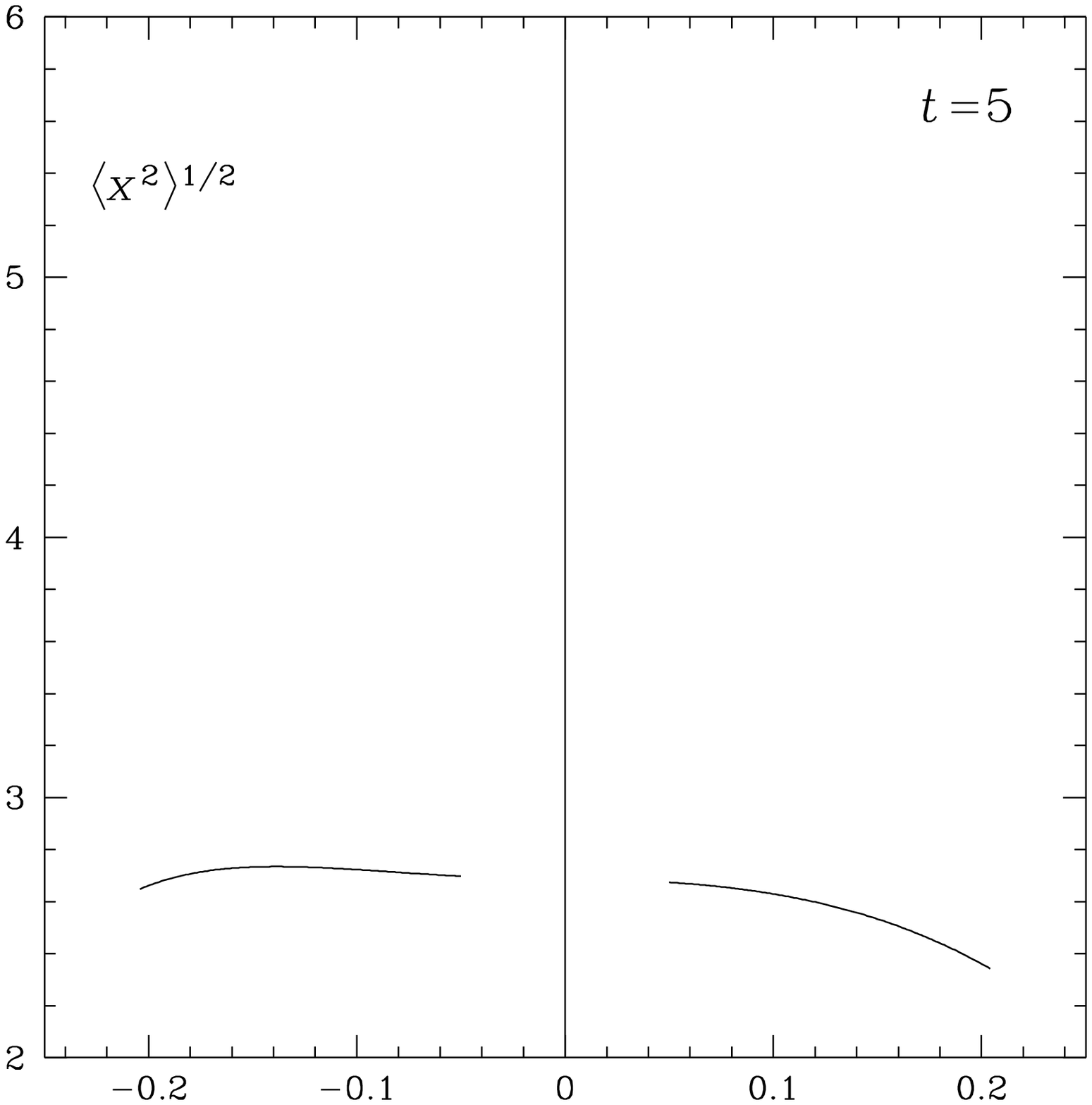}}
\centerline{\epsfxsize=3in\epsffile{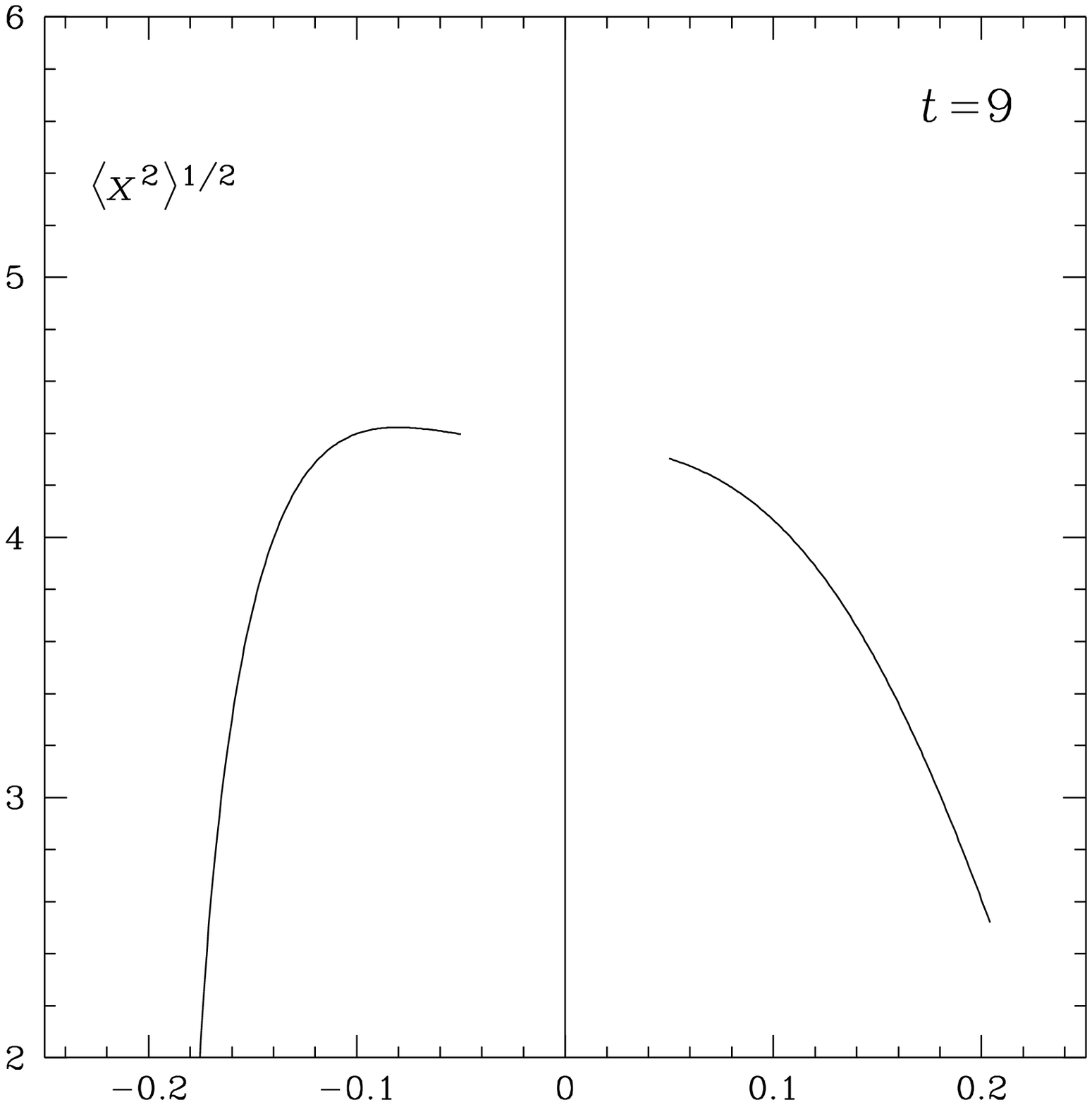}}
\centerline{\epsfxsize=3in\epsffile{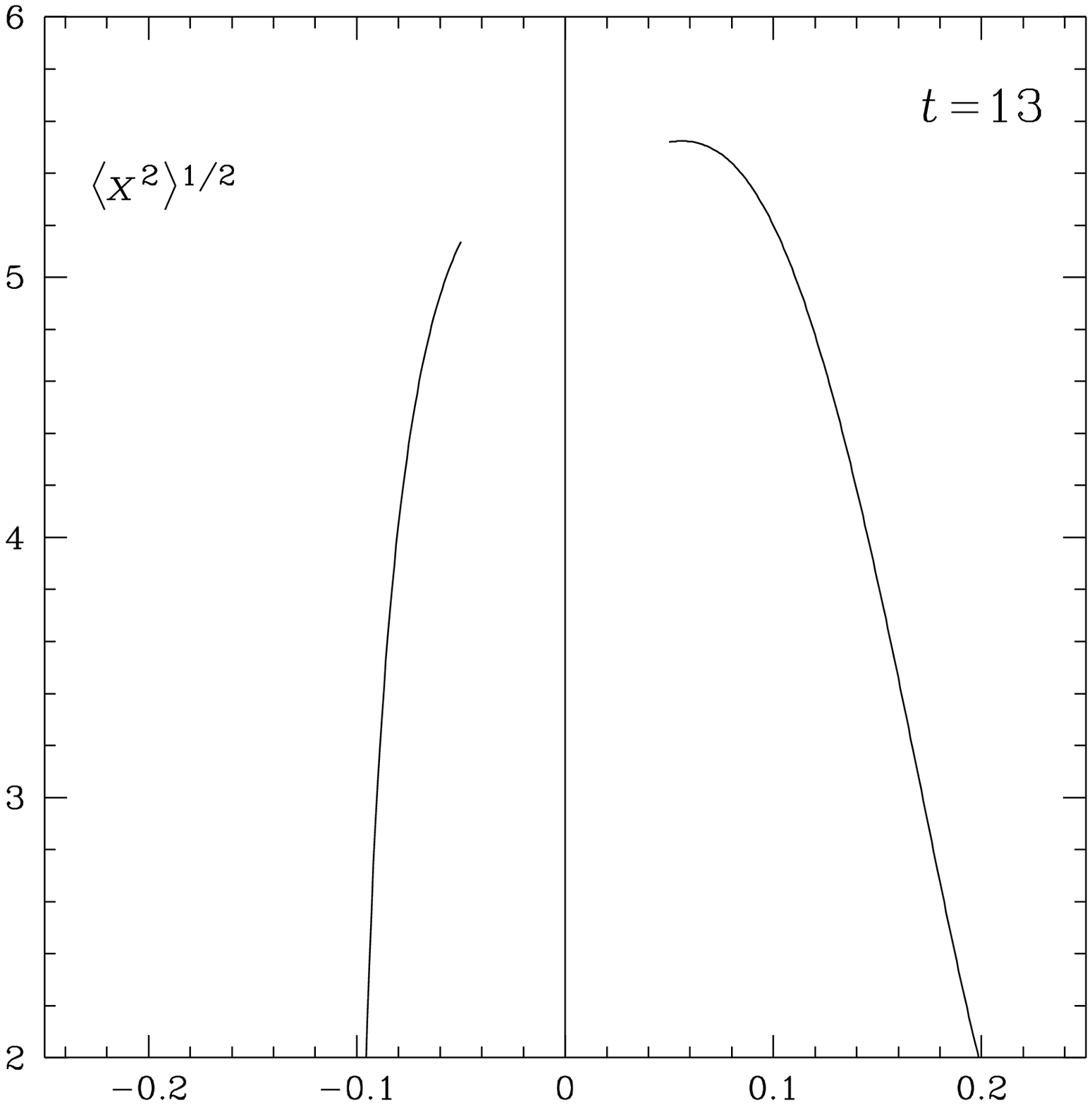}}
\caption{\label{fig1} Graphs of $\langle x^2 \rangle^\half$ versus $\sigma\mu$ for
$t$=5, 9 and 13, where $\sigma=-1$  for case (i) and $\sigma=+1$ for
case (ii). We have excluded the region $\mu<0.05$ since there
are severe round-off problems near $\mu=0$.}
\end{figure}
\newpage

\begin{figure}[t]
\vspace{-1in}
\centerline{\epsfxsize=6in \epsffile{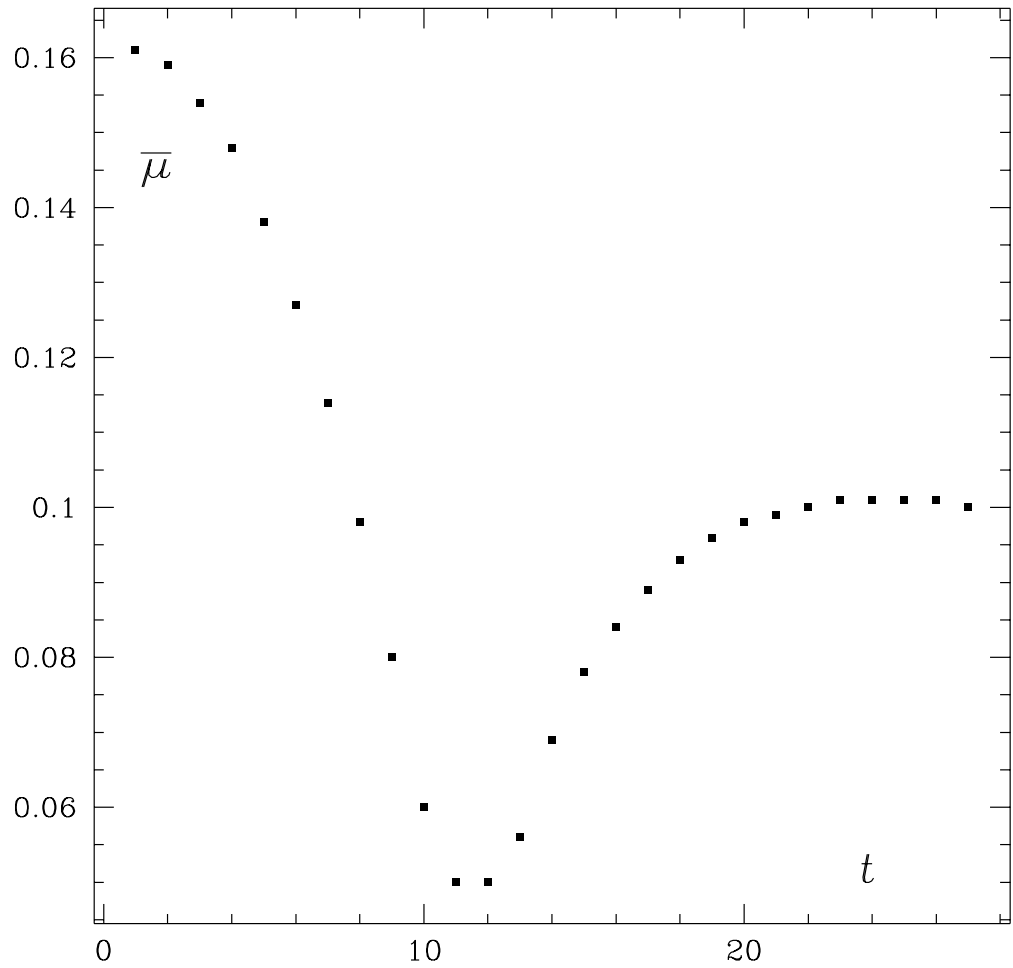}}
\vskip-0.3in
\caption{\label{fig2} $\bar\mu$ versus $t$. The change-over from case (i) to
case (ii) occurs between $t$=11 and $t$=12.}
\end{figure}
\newpage

\begin{figure}[t]
\vspace{-1in}
\centerline{\epsfxsize=8in \epsffile{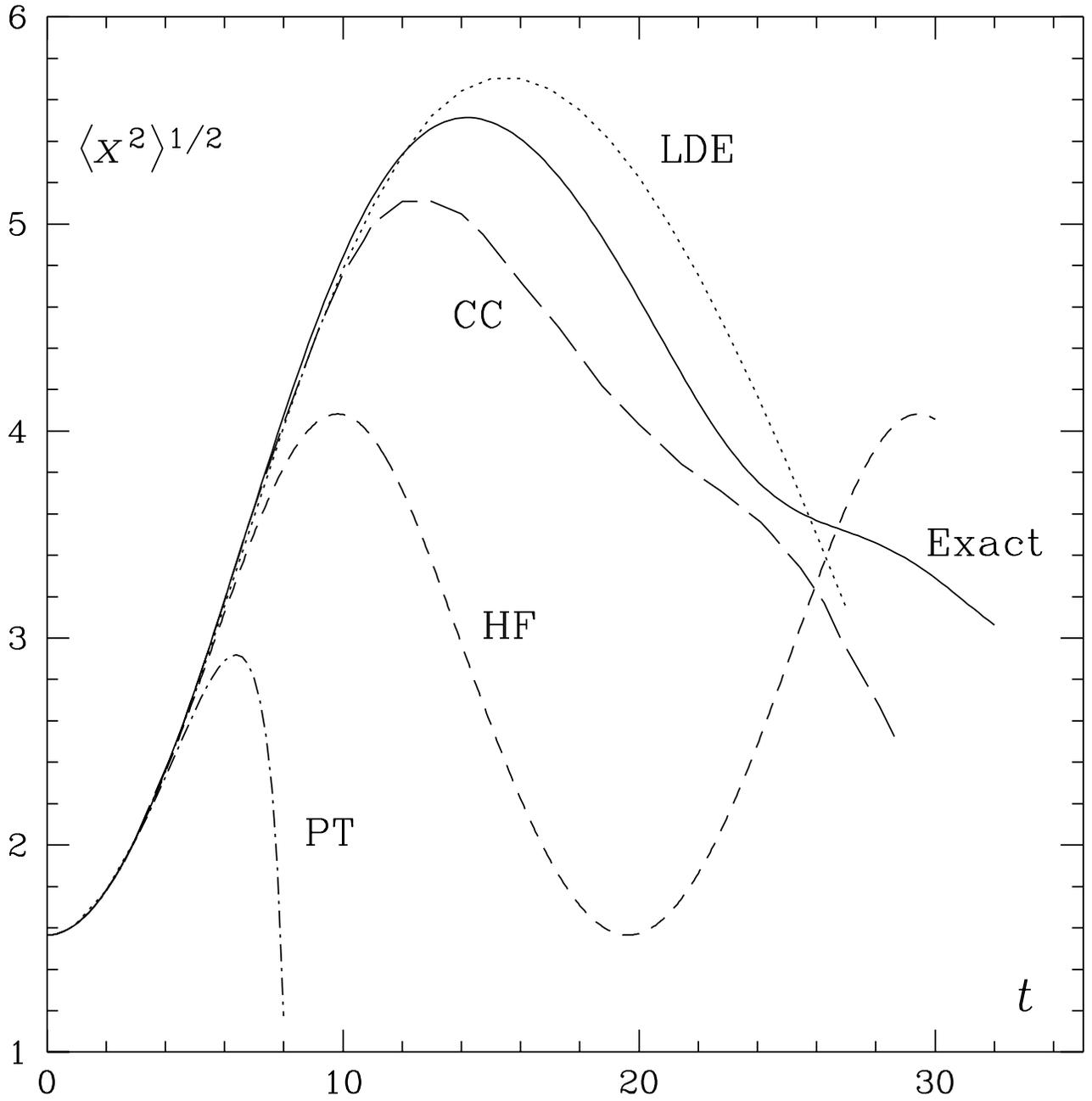}}
\caption{\label{fig3} $\langle x^2 \rangle^\half$ versus $t$.
First-order linear delta expansion
(LDE) compared with the exact result (Exact), the variational
calculations of Ref.~\protect\cite{gpi} (HF) and Ref.~\protect\cite{cpi}
(CC), and first-order
perturbation theory (PT).}
\end{figure}
\newpage

\begin{figure}[t]
\vspace{-1in}
\centerline{\epsfxsize=4in\epsffile{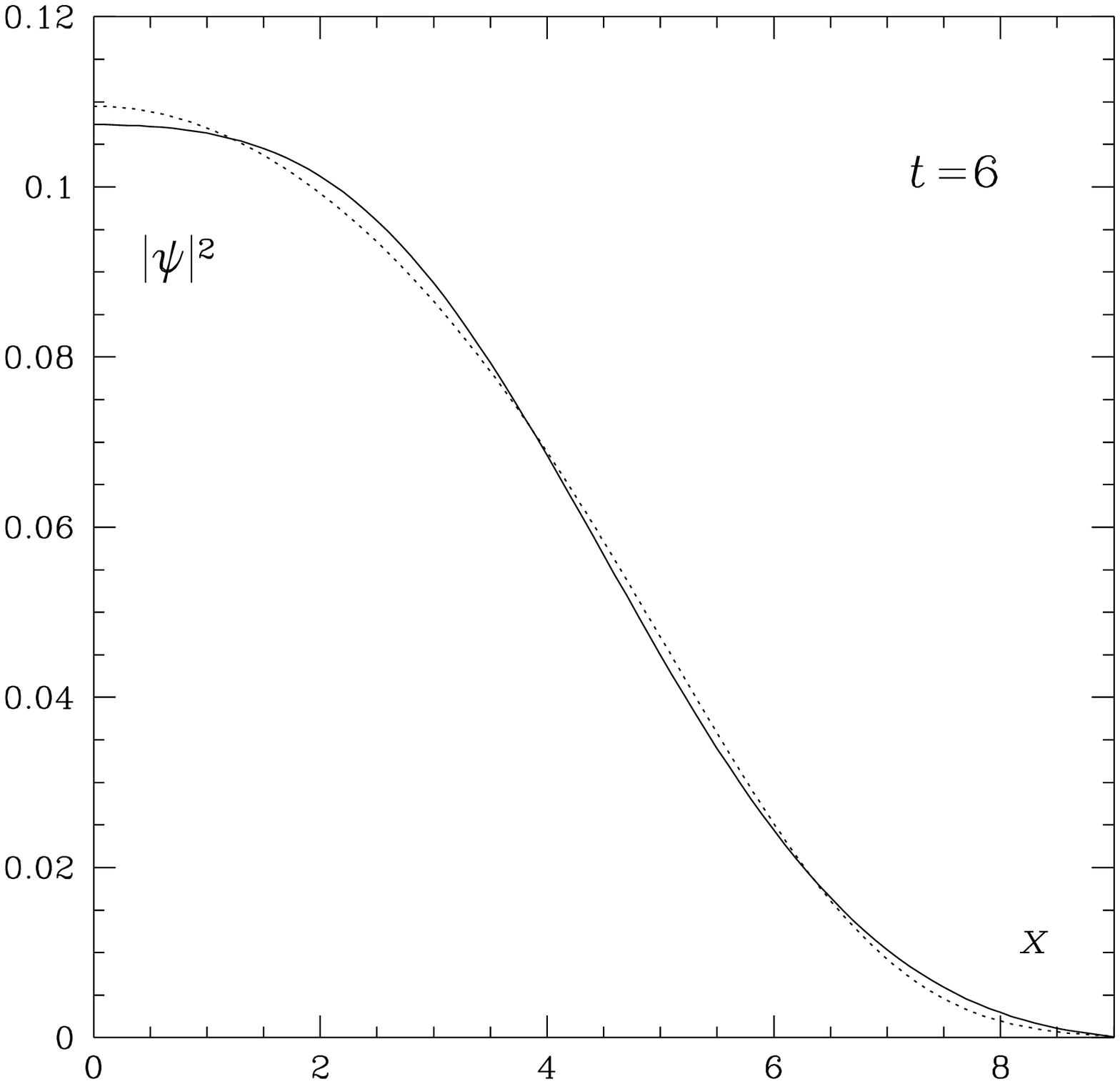}}
\centerline{\epsfxsize=4in\epsffile{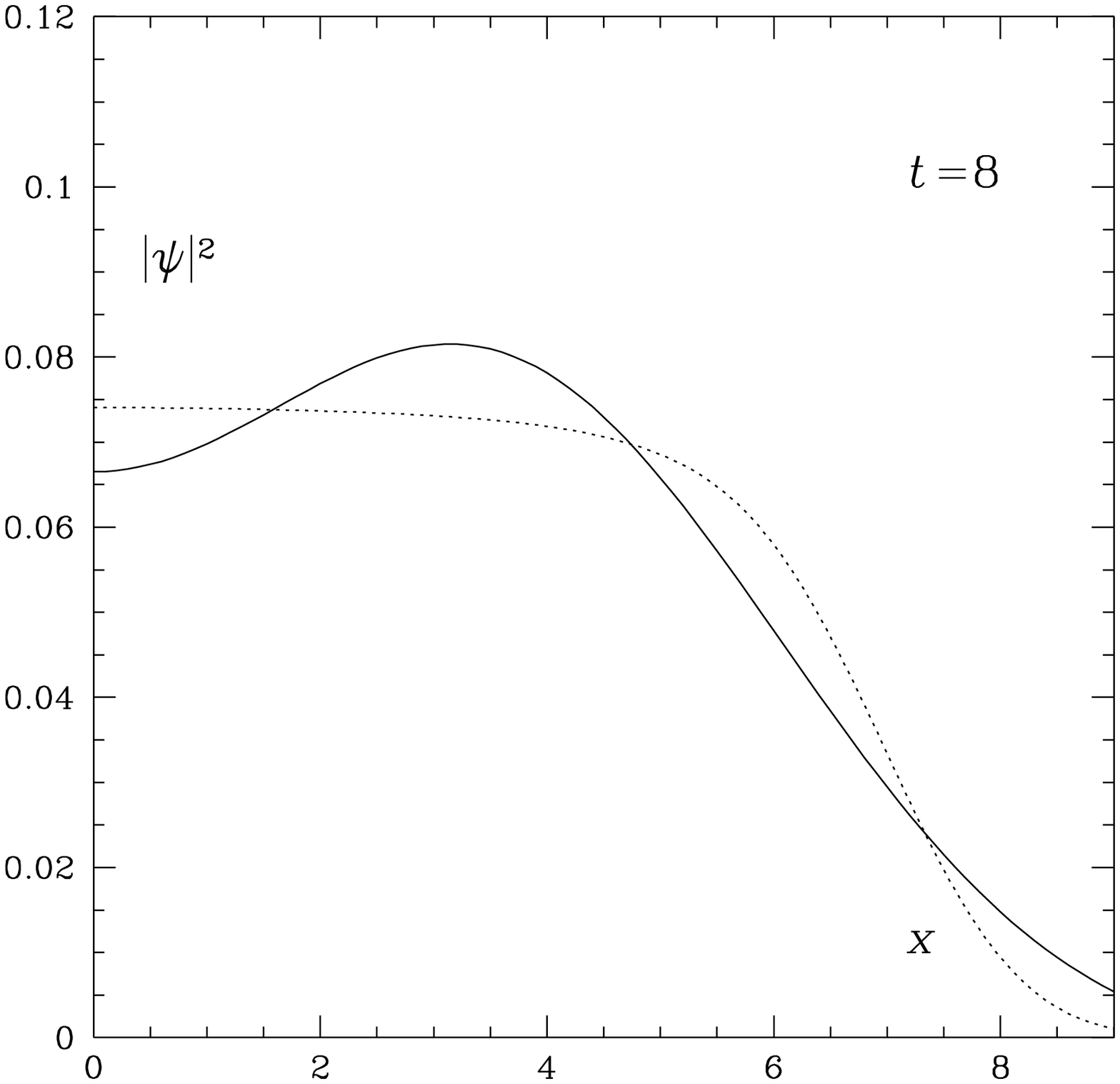}}
\caption{\label{fig4}
Graphs of $|\psi|^2$ versus $x$ for $t$=6 and 8. The
solid line is the first-order LDE calculation and the dotted line
is the exact result.}
\end{figure}


\begin{thebibliography}{99}

\bibitem{ad+hfj} {A.~Duncan and H.~F.~Jones, Phys. Rev.~{\bf D47},
2560 (1993).}

\bibitem{guida} {R.~Guida, K.~Konishi and H.~Suzuki,
Annals Phys.~{\bf 241}, 152 (1995); {\bf 249} 109 (1996).}

\bibitem{gpi} {A.~H.~Guth and S-Y.~Pi, Phys. Rev.~{\bf D32}, 1899
(1985).}

\bibitem{cpi} {F.~Cooper, S-Y.~Pi and P.~N.~Stancoff, Phys. Rev.~{\bf
D34}, 3831 (1986).}

\bibitem{cc} {G.~J.~Cheetham and E.~J.~Copeland, Phys. Rev~{\bf D53},
4125 (1996).}

\bibitem{pms} {P.~M.~Stevenson, Phys. Rev.~{\bf D23}, 2916
(1981).}

\bibitem{dm} {D.~Monteoliva, private communication.}

\bibitem{cmb} {C.~M.~Bender, K.~A.~Milton, M.~Moshe, S.~Pinsky and
L.~M.~Simmons, Jr., Phys. Rev.~{\bf D37}, 1472 (1988).}

\bibitem{fernando} {F.~C.~Lombardo, F.~D.~Mazzitelli and
D.~Monteoliva, Phys. Rev. {\bf D62}, 045016 (2000).}

\end{thebibliography}
\end{document}